\documentclass[conference,10pt]{IEEEtran}
\usepackage{diagbox} 
\usepackage{graphicx}
\usepackage{epstopdf}
\usepackage{psfrag}
\usepackage{subfigure}
\usepackage{url}
\usepackage{stfloats}
\usepackage{amsfonts,amssymb,amsmath,bm,paralist,theorem,cite,ifthen,color,nccmath}
\usepackage{caption}
\usepackage{calc}
\usepackage{enumerate}
\usepackage{multirow}
\usepackage{makecell}
\usepackage[ruled]{algorithm2e}

\usepackage{array}
\usepackage{float}
\usepackage{soul}
\usepackage[colorlinks,
            linkcolor=blue,
            anchorcolor=red,
            citecolor=red]{hyperref}

\usepackage[T1]{fontenc}
\graphicspath{{Figures/}}

\newcommand{\T}{{\scriptscriptstyle\mathsf{T}}}
\renewcommand{\H}{{\scriptscriptstyle\mathsf{H}}}

\newcommand\Ccl{\ensuremath{\mathcal{C}}}
\newcommand\Dcl{\ensuremath{\mathcal{D}}}

\newcommand\Ncl{\ensuremath{\mathcal{N}}}

\newcommand\Icl{\ensuremath{\mathcal{I}}}

\newcommand\Tcl{\ensuremath{\mathcal{T}}}

\newcommand\Vcl{\ensuremath{\mathcal{V}}}

\newcommand\Xcl{\ensuremath{\mathcal{X}}}
\newcommand\Lcl{\ensuremath{\mathcal{L}}}

\newcommand\Cs{\ensuremath{{\mathbb{C}}}}
\newcommand\Es{\ensuremath{{\mathbb{E}}}}
\newcommand\Rs{\ensuremath{{\mathbb{R}}}}

\newcommand\Pbb{\ensuremath{{\mathbb{P}}}}

\newcommand\Pb{\ensuremath{ \mathbf{P} }}

\newcommand\Xb{\ensuremath{ \mathbf{X} }}

\newcommand\Zb{\ensuremath{ \mathbf{Z} }}
\newcommand\Rb{\ensuremath{ \mathbf{R} }}

\newcommand\bb{\ensuremath{ \mathbf{b} }}

\newcommand\hb{\ensuremath{ \mathbf{h} }}

\newcommand\pb{\ensuremath{ \mathbf{p} }}
\newcommand\qb{\ensuremath{ \mathbf{q} }}

\newcommand\vb{\ensuremath{ \mathbf{v} }}

\usepackage[labelsep=period,font=footnotesize]{caption}

\usepackage{etoolbox}
\newcommand{\zerodisplayskips}{%
  \setlength{\abovedisplayskip}{3pt}%
  \setlength{\belowdisplayskip}{3pt}%
  \setlength{\abovedisplayshortskip}{3pt}%
  \setlength{\belowdisplayshortskip}{3pt}}
\appto{\normalsize}{\zerodisplayskips}
\appto{\small}{\zerodisplayskips}
\appto{\footnotesize}{\zerodisplayskips}

\usepackage{titlesec}
\titlespacing{\section}{-0.64 cm}{2pt}{2pt}
\titlespacing{\subsection}{0 cm}{2pt}{2pt}
\usepackage[normalem]{ulem}

\usepackage[all=normal,paragraphs=tight,floats=normal,mathspacing=normal,wordspacing=tight,charwidths=tight,mathdisplays=normal,leading=normal]{savetrees}
\IEEEoverridecommandlockouts
\usepackage{xcolor}

\begin{document}

\title{ Lightweight Vision-Aided Beam Tracking for Cross-Environment mmWave Communications\vspace{-5mm}}
\author{
\IEEEauthorblockN{Mengyuan Ma$^*$, Ahmed Alkhateeb$^\dagger$, Nhan Thanh Nguyen$^*$, A.~Lee~Swindlehurst$^\S$, and Markku Juntti$^*$}
\IEEEauthorblockA{$^*$Centre for Wireless Communications (CWC), University of Oulu, Finland \\
$^\dagger$School of Electrical, Computer, and Energy Engineering, Arizona State University, Tempe, USA\\
$^\S$School of Electrical Engineering \& Computer Science, University of California, Irvine, USA\\
Email: \{mengyuan.ma, nhan.nguyen, markku.juntti\}@oulu.fi;alkhateeb@asu.edu;swindle@uci.edu
}\vspace{-10mm}}

\maketitle

\begin{abstract}
Sensing-aided beam tracking is a promising approach to reduce the overhead for millimeter-wave beam management. However, real-world application remains challenging due to rapid channel variations and substantial environmental differences across deployment scenarios. Developing low-complexity sensing assisted approaches that generalize to diverse environments can alleviate the problem. With this motivation, this paper proposes a lightweight vision-aided model for cross-environment beam tracking. The task is formulated as a sequence-to-sequence classification problem, where the model jointly predicts the current and future optimal beams from past visual observations. We develop a low-complexity model based on depthwise separable convolutions and introduce hierarchical data augmentation and beam power-based label smoothing to improve robustness and generalization. Experimental results on real-world images from two geometrically distinct DeepSense~6G scenarios show that the proposed strategies consistently improve cross-environment beam prediction accuracy 
up to $84\%$ across the current and three future time slots, outperforming the state-of-the-art solution. Notably, this performance is achieved while reducing the number of model parameters and computational complexity by factors of approximately $52$ and $79$, respectively, compared with the high-capacity ResNet baseline.
\end{abstract}


\IEEEpeerreviewmaketitle

\section{Introduction}

Millimeter-wave (mmWave) systems rely on highly directional transmission to achieve high data rates, making accurate beam training and tracking essential for reliable communication. In practical deployments, however, beam tracking is challenging due to rapid channel variations, user mobility, blockage, and environmental clutter~\cite{yi2024beam}. In the context of integrated sensing and communications (ISAC), sensing-aided beam tracking has emerged as a promising use case, where contextual information from widely available sensors, such as cameras and radars, is exploited to reduce the overhead of beam management~\cite{nguyen2026knowledge,xue2024ai}.

Existing studies have demonstrated the feasibility of sensing-assisted beam prediction using position information~\cite{morais2023position}, radar~\cite{Demirhan2022Radar}, vision~\cite{jiang2022computer,ma2025attention,ma2025knowledge,imran2024environment}, and LiDAR~\cite{marasinghe2022lidar}. More recent studies leverage multiple sensing modalities to enhance beam prediction performance~\cite{ma2026knowledge,raha2025advancing,patel2024harnessing,cui2024sensing,zhu2025advancing}. Nevertheless, most existing works focus on predicting the beam for the current time slot, while ignoring model efficiency in terms of memory footage and computational complexity. These methods can also incur considerable sensing, processing, and computational overhead since beam prediction may need to be performed repeatedly over time. Long-term beam prediction can alleviate this burden by jointly predicting the beams for the current and future time slots~\cite{jiang2022computer,Jiang2024LiDAR,Luo2023millimeter,ma2025attention,ma2025knowledge,ma2026knowledge}. Despite these advances, existing sensing-aided beam tracking methods mainly evaluate prediction accuracy in one particular environment with a limited attention to lightweight deployment and robustness across environment variations. 
Motivated by this gap, this paper develops a lightweight vision-aided long-term beam tracking framework with enhanced learning strategies to improve model's generalization ability and cross-environment robustness.

Unlike our previous work that studied vision-aided beam tracking within single-environment settings~\cite{ma2025attention,ma2025knowledge}, this work targets low-complexity beam tracking under multiple environments. Specifically, we formulate vision-aided long-term beam tracking as a sequence-to-sequence (Seq2Seq) classification task, where the model jointly predicts the optimal beams for the current and future time slots from past visual observations under a predefined beam codebook. We then design a lightweight vision-aided architecture that combines depthwise separable convolutions with temporal sequence modeling, enabling efficient spatial-temporal feature learning for low-complexity beam tracking. We further jointly exploit hierarchical data augmentation and beam power-based label smoothing for improved model robustness. Simulation results on two geometrically distinct real-world DeepSense~6G scenarios \cite{alkhateeb2023deepsense} demonstrate that the proposed model achieves competitive beam prediction accuracy across the current and future time slots, outperforming the state-of-the-art (SOTA) solution \cite{ma2025attention} with substantially reduced model size and computational complexity compared with a high-capacity ResNet-based baseline. 

\section{System Model and Problem Formulation}\label{sec:system model}
 \begin{figure}[t]
	\small
	\centering	
	\includegraphics[width=0.5\textwidth]{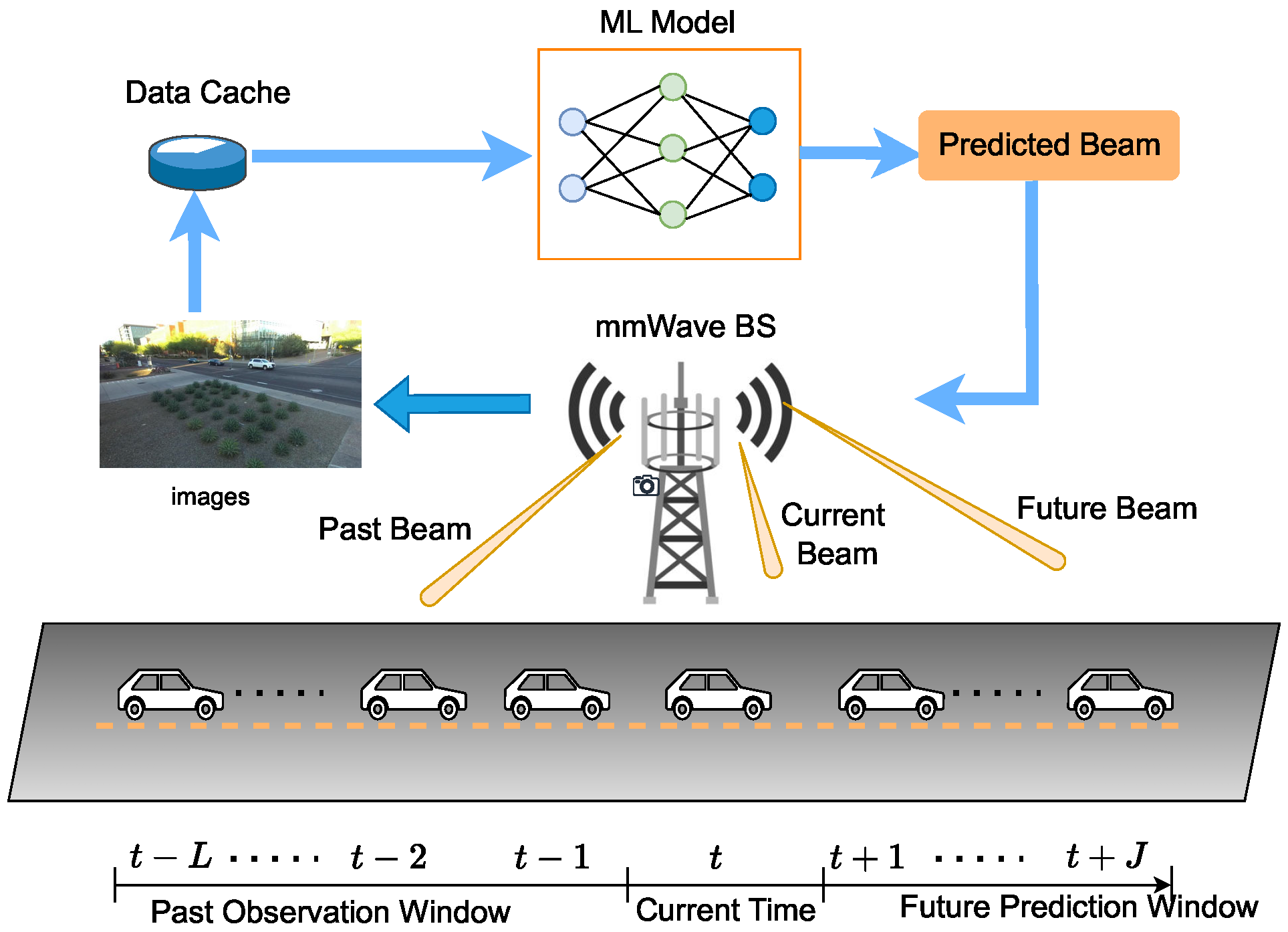}
	\vspace{-4mm}
	\caption{System illustration.}
	\label{fig:system model}
\end{figure}
We consider downlink mmWave communications aided by beam tracking and operating in different environments. A base station (BS) equipped with a uniform linear array serves a single-antenna mobile user equipment (UE), as illustrated in Fig.~\ref{fig:system model}. In addition to the communications array, the BS is equipped with an RGB camera to capture real-time UE and environmental dynamics. The camera-based system should operate effectively in different physical scenarios, which is challenging since image observations vary significantly with background, road geometry, and traffic conditions. 


\subsection{System Model}\label{sec:system model_A}
Let $i\in\{1,\ldots,I\}$ be the scenario index. At time slot $t$, the BS in scenario $i$ transmits a symbol $s_i[t]\in\Cs$ with $\Es[|s_i[t]|^2]=1$ to the UE. Let $\vb_i[t]\in\Vcl$ denote the beamforming vector selected from a predefined beam codebook $\Vcl$. We assume that the same codebook is used across all $I$ scenarios. The received signal for the UE at the $i$-th scenario is given by
\begin{align}\label{eq:signal_model}
y_i[t] = \hb_i[t]^\H \vb_i[t] s_i[t] + n_i[t],
\end{align}
where $\hb_i[t]$ denotes the channel in scenario $i$, and $n_i[t]\sim\Ccl\Ncl(0,\sigma_{\rm n}^2)$ models noise and other interference. The corresponding SNR is \( \mathrm{SNR}_i[t]=\frac{|\hb_i[t]^\H\vb_i[t]|^2}{\sigma_{\rm n}^2} \).


\subsection{Problem Formulation}
At time slot $t$, our goal is to determine the transmit beamforming vectors at the BS for the current and $J$ future time slots $\{t,t+1,\ldots,t+J\}$ over $I$ scenarios. Let $\Vcl=\{\vb_1,\ldots,\vb_C\}$ and $\Icl_{\Vcl}=\{1,\ldots,C\}$ denote the beamforming codebook and its associated index set, respectively, where $C\triangleq|\Vcl|$. In the considered multi-scenario beam tracking problem, we aim to find $\vb_i[\tau]\in\Vcl$, $\forall i,\tau$, to maximize the accumulated spectral efficiency $    R_J = \sum_{i=1}^{I}\sum_{\tau=t}^{t+J}
    \log\left(1+\mathrm{SNR}_i[\tau]\right)$
over all scenarios and time slots. This beamforming problem can be reformulated as~\cite{Jiang2024LiDAR,Luo2023millimeter,ma2025attention,ma2025knowledge}
\begin{equation}\label{pb:P2}
    \underset{\vb_i[\tau]\in\Vcl,\ \forall i,\tau}{\rm maximize}
    \quad
    \sum_{i=1}^{I}\sum_{\tau=t}^{t+J}
    \left|\hb_i[\tau]^\H \vb_i[\tau]\right|^2 .
\end{equation}
Let $\qb_i^{\star}[t]=\big[q_i^{\star}[t],\ldots,q_i^{\star}[t+J]\big]^\T$ denote the vector of optimal beam indices for scenario $i$. Then, the corresponding optimal beam-index sequence is given by
\begin{equation}\label{pb:P3}
    \left\{\qb_i^{\star}[t]\right\}_{i=1}^{I}
    =
    \underset{q_i[\tau]\in\Icl_{\Vcl},\ \forall i,\tau}{\arg\max}
    \quad
    \sum_{i=1}^{I}\sum_{\tau=t}^{t+J}
    \left|\hb_i[\tau]^\H \vb_{q_i[\tau]}\right|^2 .
\end{equation}

One solution to \eqref{pb:P3} is to decouple it into $I(J+1)$ independent subproblems, where each subproblem is solved by exhaustive search over the $C$ candidate beams or by training an ML model to map sensory observations to the optimal beam~\cite{Demirhan2022Radar,morais2023position,imran2024environment,marasinghe2022lidar,cui2024sensing,zhu2025advancing}. However, exhaustive search requires perfect channel state information (CSI), while per-slot ML-based prediction requires frequent inference across time slots and scenarios, making both approaches inefficient for long-term beam tracking.

An alternative is to train a sequential ML model that exploits temporal dependencies in sensory observations to jointly predict the current and future beams~\cite{Jiang2024LiDAR,Luo2023millimeter,ma2025attention,ma2025knowledge,ma2026knowledge}. While effective, existing approaches are typically limited to the single-scenario case with $I=1$, and their robustness for $I>1$ remains insufficiently explored. This work addresses this gap by developing a lightweight vision-aided beam tracking framework with enhanced learning strategies for robust cross-environment prediction.

\section{Vision-Aided Lightweight Beam Tracking}
This section presents the proposed framework for long-term beam tracking from visual observations. We first formulate the beam tracking problem as a Seq2Seq classification task and then introduce the lightweight architecture designed for efficient spatial-temporal feature learning. Details are provided below.

\subsection{Learning Task}
Let $\Xb_i[t]\in\Rs^{d_{\rm C}\times d_{\rm H}\times d_{\rm W}}$ denote the RGB image obtained at time slot $t$ in scenario $i$, where $d_{\rm C}$, $d_{\rm H}$, and $d_{\rm W}$ denote the number of RGB color channels, the image height, and the image width, respectively. Define 
$\Xcl_i[t]=\{\Xb_i[t-L],\Xb_i[t-L+1], \ldots,\Xb_i[t]\}$ as the sequence of raw visual observations from the past $L$ time slots to the current time slot $t$ in scenario $i$. 

We aim to develop an efficient learning model to solve \eqref{pb:P3}, i.e., to predict the optimal beam indices in $\Icl_{\Vcl}$ for the current and future time slots $\{t,t+1,\ldots,t+J\}$ across the considered scenarios. Let $f(g(\Xcl_i[t]);\Theta)$ denote the learning model with trainable parameters $\Theta$, where $g(\cdot)$ represents the preprocessing operations for raw images. The model outputs the beam selection probabilities for all $C$ candidate beams over the $J+1$ time slots. Let $p_{i,c}[t+j]$ denote the probability of selecting the $c$-th beam in scenario $i$ at time slot $t+j$, and define $\pb_i[t+j]=[p_{i,1}[t+j],\ldots,p_{i,C}[t+j]]^\T\in\Rs^{C}$.
The predicted beam index is obtained as
\begin{equation}\label{eq:predicted beam index}
    \hat{q}_i[\tau]
    =
    \arg\max_{c\in\Icl_{\Vcl}} \; p_{i,c}[\tau],
    \quad \forall i, \tau=t,\ldots,t+J .
\end{equation}
The desired learning model for vision-aided beam tracking is written as
\begin{equation}\label{pb:ML task}
    \Theta^{\star}
    =
    \arg\max_{\Theta}
    \quad
    \sum_{i=1}^{I}\sum_{\tau=t}^{t+J}
    \Pbb\{\hat{q}_i[\tau]=q_i^{\star}[\tau]\},
\end{equation}
where $\Pbb\{\cdot\}$ denotes an event probability. We note that $L$ and $J$ are empirically determined hyperparameters.

\subsection{Lightweight Beam Tracking ML Model Structure}\label{sec:model structures}

\begin{figure}[t]
\small
\hspace{-2mm}
  \includegraphics[width=0.5\textwidth]{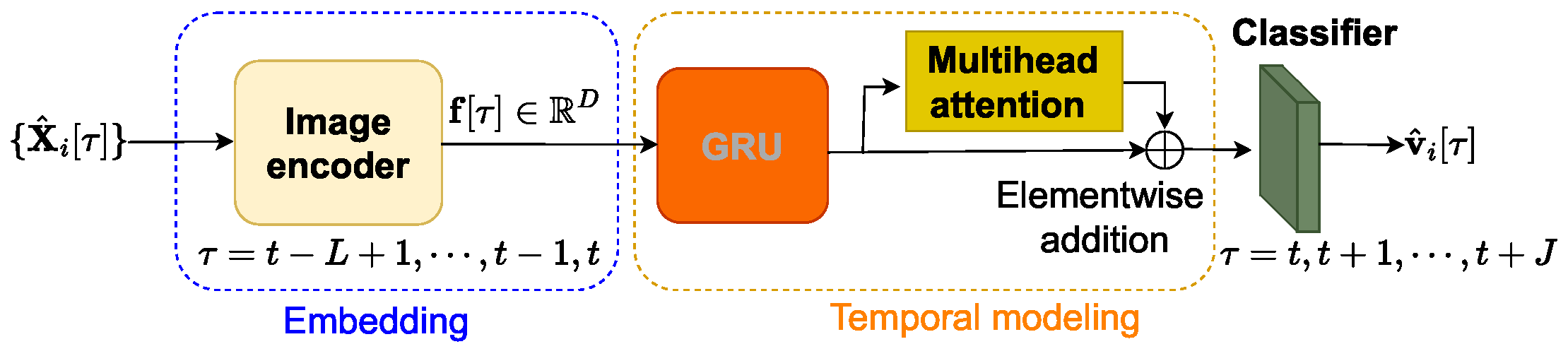}
\vspace{-5mm}
\caption{Proposed lightweight vision-aided model structure for beam tracking.}
\label{fig:Model structure}
\end{figure}

As shown in \eqref{pb:ML task}, the considered learning task is to predict the beam selections over the current and future time slots from a sequence of past vision observations. 
We develop a lightweight framework based on efficient image embedding, temporal modeling, and beam classification, as illustrated in Fig.~\ref{fig:Model structure}. The temporal modeling consists of a gated recurrent unit (GRU) network and a multihead attention (MHA) module, which can effectively capture the time dependency of sequential data~\cite{ma2025attention,ma2025knowledge}.

\begin{figure}[t]
\small
\centering	
  \includegraphics[width=0.48\textwidth]{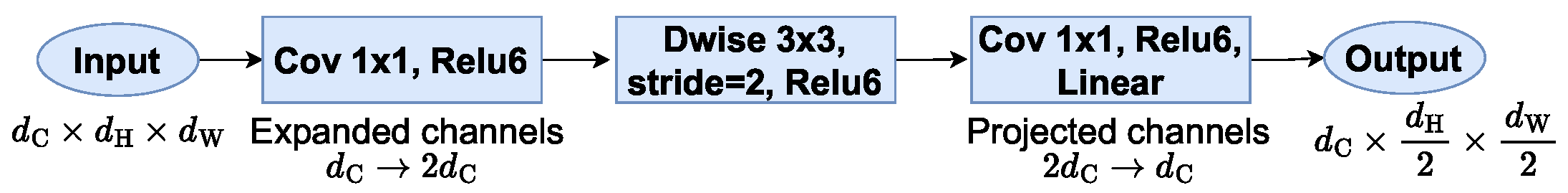}
\vspace{-2mm}
\caption{Designed DS-Conv block.}
\label{fig:DS-Cov_block}
\end{figure}

The image encoder accounts for most of the complexity in the proposed architecture. To reduce this burden, we adopt a lightweight encoder consisting of a ResNet-style stem followed by depthwise separable convolution (DS-Conv) blocks~\cite{sandler2018mobilenetv2}. The overall structure is summarized in Table~\ref{tab:lightweight_structure}. The convolutional stem uses a relatively large $7\times 7$ kernel to extract low-level spatial features. Each subsequent DS-Conv module is implemented as a MobileNetV2-style inverted residual downsampling block, where a $1\times 1$ pointwise convolution first expands the channel dimension, a $3\times 3$ depthwise (Dwise) convolution with stride 2 performs spatial downsampling, and a final $1\times 1$ pointwise linear convolution projects the features to the target channel dimension, as illustrated in Fig.~\ref{fig:DS-Cov_block}. Since each block changes both the spatial resolution and the number of channels, the residual shortcut is omitted. This design improves feature representation by first expanding and then compressing channels of the feature maps while keeping the encoder lightweight. After four DS-Conv blocks, global max pooling is applied to compress the feature maps. The resulting feature sequence is then processed by a single-layer GRU and an MHA module for temporal modeling, followed by a three-layer multilayer perceptron (MLP) classifier for joint current-and-future beam prediction.


The lightweight design reduces deployment complexity but may limit the model's robustness under cross-environment deployment. We therefore introduce an enhanced learning strategy to improve its generalization ability.

\begin{table}[t]
\centering
\caption{Compact architecture of the proposed lightweight model.}
\label{tab:lightweight_structure}
\renewcommand{\arraystretch}{1.1}
\setlength{\tabcolsep}{5pt}
\begin{tabular}{ll}
\hline
\textbf{Stage} & \textbf{Structure} \\
\hline
{\bf Input} & Preprocessed image  sequence \\
\hline
\multicolumn{2}{l}{\textbf{Image encoder}} \\
Stem & Conv$(7\times7,16)$ + BN + ReLU + MaxPool \\
Block 1 & DS-Conv$(16\!\rightarrow\!32)$ \\
Block 2 & DS-Conv$(32\!\rightarrow\!64)$ \\
Block 3 & DS-Conv$(64\!\rightarrow\!128)$ \\
Block 4 & DS-Conv$(128\!\rightarrow\!256)$ \\
Global Pooling &  AdaptiveMaxPool2d \\
\hline
\textbf{Temporal modeling} & LayerNorm + 1-layer GRU + MHA \\
\textbf{Classifier} & 3-layer MLP, output dimension $(J+1)\times C$ \\
\hline
\end{tabular}
\end{table}

\section{Cross-Scenario Lightweight Learning }\label{sec:learning}
We first convert the raw visual observations into task-oriented representations using adjacent-frame subtraction to suppress background interference and extract binary motion masks that highlight the moving UE~\cite{ma2025knowledge}. These preprocessed representations are then fed into the image encoder. Building on this input representation, we improve the learning efficiency by designed hierarchical data augmentation and beam power-based label smoothing.

\subsection{Hierarchical Data Augmentation}
Data augmentation improves model robustness by encouraging invariant feature learning from diverse data variations~\cite{zhang2017mixup}. For vision-aided beam tracking, we design physically meaningful augmentations to emulate imperfect sensing conditions while preserving the temporal motion patterns of the UE. Since the input is a sequence of image observations, the proposed hierarchical augmentation operates at both the sequence and frame levels, as detailed below.

Let $\Xb^{\mathrm{batch}} \in \mathbb{R}^{B \times L \times d_{\rm C} \times d_{\rm H} \times d_{\rm W}}$ denote a batch of image sequences, where $B$ is the batch size. For each sample\footnote{We note that one sample corresponds to one observation sequence $\Xcl[t]$ while one observation $\Xb[t]$ is referred to a frame. We omit the scenario index $i$ in the sequel for notational simplicity.}, we apply sequence-consistent spatial transformations with certain probabilities, followed by frame-wise additive noise. Specifically, to account for imperfections in motion-mask extraction, we apply a sequence-level morphological perturbation to the vision input. For each image sequence, we first sample a binary variable \(z_b^{\mathrm{morph}}\sim\mathrm{Bernoulli}(p_{\mathrm{morph}}), \) with probability $p_{\mathrm{morph}}$. If $z_b^{\mathrm{morph}}=1$, one operation $o_b$ is uniformly sampled from $\{\mathrm{erosion},\mathrm{dilation}\}$. The same operation is then applied to all frames in the selected sequence:
\begin{equation}\label{eq:morph perturbation}
    \widetilde{\Xb}^{\mathrm{batch}}_{b,\tau}=
\mathcal{M}_{o_b}^{(3\times3)}\left(\Xb^{\mathrm{batch}}_{b,\tau}\right),
\ \tau=t,t-1,\ldots,t-L,
\end{equation}
where $\mathcal{M}_{o_b}^{(3\times3)}(\cdot)$ denotes the selected morphological operation using a $3\times3$ structuring kernel. Here, erosion slightly shrinks the foreground region, whereas dilation slightly expands it. These operations emulate under- and over-segmentation caused by imperfect background subtraction or thresholding. Applying the same operation across all frames preserves temporal consistency and encourages the model to be robust to small boundary errors in the extracted motion masks.

In addition, Gaussian blur is applied to emulate degraded visual observations caused by sensing uncertainty or imperfect preprocessing. Similar to morphological perturbations, each sequence is independently selected by \(z_b^{\mathrm{blur}}\sim\mathrm{Bernoulli}(p_{\mathrm{blur}})\) with probability $p_{\mathrm{blur}}$. For a selected sequence, the blur strength is sampled as $\sigma_b\sim\mathcal{U}(0.5,2.0)$, and the same Gaussian filter $\mathcal{G}_{\sigma_b}(\cdot)$ is applied to every frame:
\begin{equation}\label{eq:Gaussian blur}
    \widehat{\Xb}_{b,\tau}^{\rm batch}=\mathcal{G}_{\sigma_b}(\widetilde{\Xb}_{b,\tau}^{\rm batch}), \ \tau=t,t-1,\ldots,t-L.
\end{equation}
These sequence-level augmentations emulate realistic visual variations while preserving the underlying temporal motion pattern.

After the sequence-level perturbations, frame-wise additive Gaussian noise is further introduced to emulate local appearance fluctuations. Specifically, for each frame $\widehat{\mathbf{X}}_{b,\tau}^{\rm batch}$, we sample $z_{b,\tau}^{\mathrm{noise}}\sim \mathrm{Bernoulli}(p_{\mathrm{noise}})$. If $z_{b,\tau}^{\mathrm{noise}}=1$, additive Gaussian noise is applied as
\begin{equation}\label{eq:vision AWGN}
    \overline{\Xb}_{b,\tau}^{\rm batch}= \mathrm{clip}_{[0,1]}\left(\widehat{\Xb}_{b,\tau}^{\rm batch}+\mathbf{N}_{b,\tau}\right),
\end{equation}
where $\mathbf{N}_{b,\tau}\sim \mathcal{N}\left(\mathbf{0},\sigma_{\rm image}^2\mathbf{I}\right)$ has the same dimension as $\widehat{\mathbf{X}}_{b,\tau}^{\rm batch}$. If $z_{b,\tau}^{\mathrm{noise}}=0$, the frame is kept unchanged, i.e. $\overline{\mathbf{X}}_{b,\tau}^{\rm batch}
=
\widehat{\mathbf{X}}_{b,\tau}^{\rm batch}$. This frame-wise noise injection improves robustness to small local intensity variations. 

\subsection{Beam Power-Based Label Smoothing}
Label smoothing is an effective regularization technique for mitigating over-confident predictions and improving generalization~\cite{szegedy2016rethinking}. In beam prediction, several beam classes may provide comparable beamforming gains and communication performance despite having distinct labels. Thus, strict one-hot supervision can be overly restrictive and unable to reflect the similarity among near-optimal beams. To alleviate this issue, we construct a softened beam-power distribution for each frame and use it as auxiliary supervision. With $\Lcl_{\rm hard}$ and $\Lcl_{\rm soft}$ denoting the hard- and soft-label losses, respectively, the overall loss function is given by
\begin{equation}\label{eq:overall_loss}
\mathcal{L}_{\rm}
=(1-\lambda)\mathcal{L}_{\mathrm{hard}}+\lambda \mathcal{L}_{\mathrm{soft}},
\end{equation}
where $\lambda\in[0,0.5)$ controls the contribution of the soft-label supervision. We use the Focal loss for hard-label and cross-entropy loss for the soft-label, respectively.

{\bf Hard-label loss:} Let \(\Dcl_{\rm}=\{\Xcl[t], \bb^\star [t],\Pb^\star[t]\}, t=0,\ldots, T_{\rm D}\}\) be the set of data sequences from the source dataset, where $T_{\rm D}$ denotes the number of time slots over which vision data are collected. Here, $\Xcl[t]$ represent the raw image sequence containing the past $L$ frames, $\bb^\star [t]$ and $\Pb^\star[t]=[\pb^\star[t],\ldots,\pb^\star[t+J]]\in \Rb^{C\times (J+1)}$ collects the corresponding optimal beam indices and the beam power vectors for the current and $J$ future time slots. Let $\mathbf{z}[\tau]=[z_{1},\ldots,z_{C}]\in\mathbb{R}^{C}$ be the output logit vector for the frame $\Xb[\tau]\in \Xcl[t]$. The predicted posterior probability of beam class $c$ is therefore \(\pi_{c}[\tau]=\frac{\exp(z_{c})}{\sum_{j=1}^{C}\exp(z_{j})}\).
The Focal loss is given by
\begin{equation}\label{eq:Focal loss}
    l_{\rm Focal}[\tau]=-\alpha(1-\pi_{b^{\star}}[\tau])^{\gamma}\log\left(\pi_{b^{\star}}[\tau] \right),
\end{equation}
where $\pi_{b^{\star}}[\tau]$ denotes the predicted probability of selecting the ground-truth beam index $b^{\star}[\tau]$ at time slot $\tau$. The hyperparameter $\alpha$ is the weighting factor addressing class imbalance, and $\gamma$ is the focusing parameter that de-emphasizes easy examples. The overall Focal loss for the sequence $\Xcl[t]$ is expressed as
\begin{equation}\label{eq:hard label}
    \mathcal{L}_{\mathrm{hard}}[t] =\sum_{\tau=t}^{t+J} l_{\rm Focal}[\tau].
\end{equation}
    
{\bf Soft-label loss:} Based on the beam power vector \(\pb^\star[\tau]=[p_{1}^\star,\ldots,p_{C}^\star]^\top\), we construct the soft target distribution $\hat{p}_{c}^\star
=\frac{(p_{c}^\star+\epsilon)^{1/T}}
{\sum_{j=1}^{C}(p_{j}^\star+\epsilon)^{1/T}}, c=1,\ldots,C,$
where $\epsilon>0$ is a small constant for numerical stability and $T>0$ is the temperature parameter. A smaller $T$ produces a sharper distribution that places more probability mass on high-power beams, while a larger $T$ yields a smoother distribution approaching the uniform distribution. The soft-label cross-entropy loss for the frame $\Xb[\tau]$ is written as
\begin{equation}\label{eq:soft label}
l_{\mathrm{soft}}[\tau]
=-\sum_{c=1}^{C}\hat{p}_{c}^\star\log p_{c}[\tau],
\end{equation}
where $p_{c}[\tau]$ is the predicted probability of candidate beam~$c$ at time $\tau$. The overall soft-label loss for the sequence $\Xcl[t]$ is thus
\begin{equation}
    \Lcl_{\rm soft}[t]=\sum_{\tau=t}^{t+J} l_{\rm soft}[\tau]. 
\end{equation}
Exploiting the soft label makes the model less prone to becoming overly confident on a single beam class, which is beneficial for improving robustness and generalization in cross-environment beam tracking, as will be verified in Section~\ref{sec:simulation}.

\subsection{Learning Procedure}
\begin{algorithm}[t]
\small
\caption{Enhanced Learning Procedure for Solving Problem~\eqref{pb:ML task}.}
\label{alg1}
\LinesNumbered
\KwIn{Training set $\Dcl_{\rm tr}$, validation set $\Dcl_{\rm evl}$}
\KwOut{Optimized model parameters $\Theta^{\star}$}

Initialize $\Theta$, $\Theta^{\star}$, and $\Lcl_{\rm evl}^{\star}$\;

\For{$e=1,\ldots,E$}{
    \parbox[t]{0.92\linewidth}{Randomly divide $\Dcl_{\rm tr}$ into mini-batches $\{\Dcl_{\rm tr}^{(n)}\}_{n=1}^{N_{\rm b}}$ \\ with batch size $B$\;}
    
    \For{$n=1,\ldots,N_{\rm b}$}{
    \parbox[t]{0.92\linewidth}{Perform data preprocessing \\ $\Xb^{\rm batch}=g\left(\left\{\Xcl[t],t\in \Tcl^{(n)}=\{t_1^{(n)},\ldots,t_B^{(n)}\}\right\}\right)$\;}

        \parbox[t]{0.92\linewidth}{
         Obtain augmented vision data $\overline{\Xb}^{\rm batch}$ using \eqref{eq:morph perturbation}, \eqref{eq:Gaussian blur}, \\and \eqref{eq:vision AWGN}\; }

        Obtain prediction logits $\Zb=f\left(\overline{\Xb}^{\rm batch};\Theta\right)$\;

        \parbox[t]{0.92\linewidth}{Compute average hard loss over the batch in $J+1$ \\ time slots:
        $\Lcl_{\rm hard}=\frac{1}{B(J+1)}\sum_{t\in \Tcl^{(n)}}\Lcl_{\rm hard}[t]$\;}

       \parbox[t]{0.92\linewidth}{Compute average soft loss over the batch in $J+1$ \\time slots:
        $\Lcl_{\rm soft}=\frac{1}{B(J+1)}\sum_{t\in \Tcl^{(n)}}\Lcl_{\rm soft}[t]$\;}
        
\parbox[t]{0.92\linewidth}{Obtain the overall loss in \eqref{eq:overall_loss} and \\ update $\Theta$ with an optimizer\;}
    }
    
    \parbox[t]{0.92\linewidth}{Evaluate the model on $\Dcl_{\rm evl}$ without augmentation and \\obtain $\Lcl_{\rm evl}^{(e)}$\;}
    
    \If{$\Lcl_{\rm evl}^{(e)} < \Lcl_{\rm evl}^{\star}$}{
        $\Theta^{\star}=\Theta$, $\Lcl_{\rm evl}^{\star}=\Lcl_{\rm evl}^{(e)}$\;
    }
}
\Return{$\Theta^{\star}$}\;
\end{algorithm}

Algorithm~\ref{alg1} summarizes the training procedure for the proposed beam prediction model. The training and validation datasets are denoted by $\Dcl_{\rm tr}$ and $\Dcl_{\rm evl}$, respectively. First, the model parameters $\Theta$, the best model parameters $\Theta^\star$, and the best validation loss $\Lcl_{\rm evl}^{\star}$ are initialized. For each training epoch, $\Dcl_{\rm tr}$ is randomly partitioned into mini-batches in step~3. The model is then optimized in a batch-wise manner, as shown in steps~4--11. For each mini-batch, the raw image observations are first preprocessed through $\Xb^{\rm batch}=g\left(\left\{\Xcl[t],t\in \Tcl^{(n)}=\{t_1^{(n)},\ldots,t_B^{(n)}\}\right\}\right)$ in step~5. The preprocessed image batch is subsequently augmented in step~6 using the proposed vision-domain augmentation operations, including morphological perturbation, Gaussian blur, and additive Gaussian noise. The augmented images are fed into the network in step~7 to obtain the prediction logits. Based on these logits, the hard-label and soft-label losses are computed over the mini-batch and all $J+1$ prediction time slots in steps~8 and~9, respectively. The overall loss is then obtained according to \eqref{eq:overall_loss}, and the model parameters are updated using an optimizer in step~10. After each epoch, the model is evaluated on the validation set $\Dcl_{\rm evl}$ without data augmentation in step~12. If the validation loss decreases, the best model parameters $\Theta^\star$ and the corresponding validation loss $\Lcl_{\rm evl}^{\star}$ are updated in steps~13--15. The optimal model parameters $\Theta^{\star}$ are returned when the maximum number of epochs is reached or the best validation loss $L_{\rm evl}^{\star}$ stops decreasing over a predefined number of consecutive epochs.


\section{Numerical Results and Discussion}\label{sec:simulation}
\begin{figure}[t]
\vspace{-2mm}
\small
    \centering
        \subfigure[Scenario 9.]
    {\label{fig:S9_layout} \includegraphics[width=0.235\textwidth]{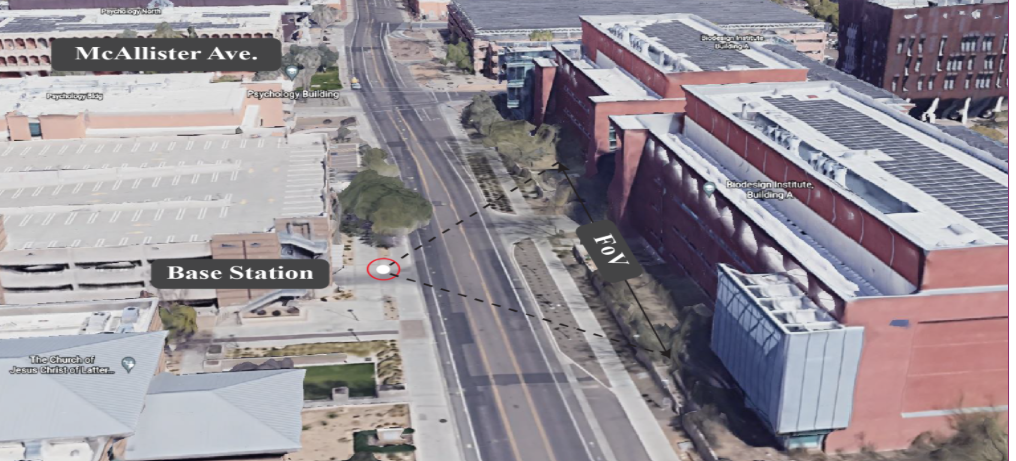}}
        \subfigure[Scenario 31.]
    {\label{fig:S31_layout} \includegraphics[width=0.235\textwidth]{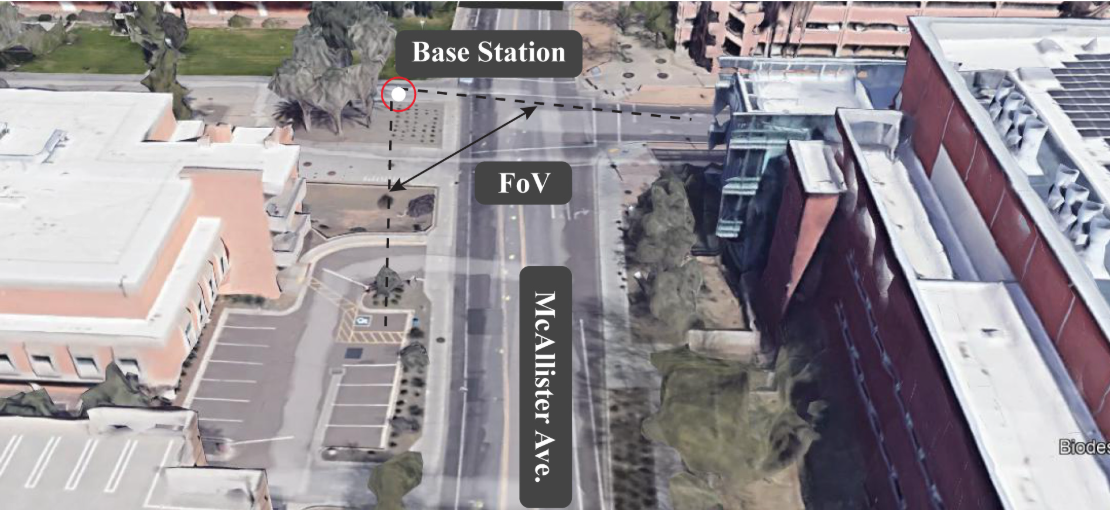}}
    \vspace{-2mm}
    \caption{Layout illustrations of DeepSense 6G scenarios 9 and 31 \cite{alkhateeb2023deepsense}, showing the BS location, camera field of view (FoV), and surrounding road environment. }
    \label{fig:sceanrio_illustration}
\end{figure}

\begin{figure}[t]
\vspace{-3mm}
\small
    \centering
        \subfigure[ Scenario 9.]
    {\label{fig:S9_beam_density} \includegraphics[width=0.235\textwidth]{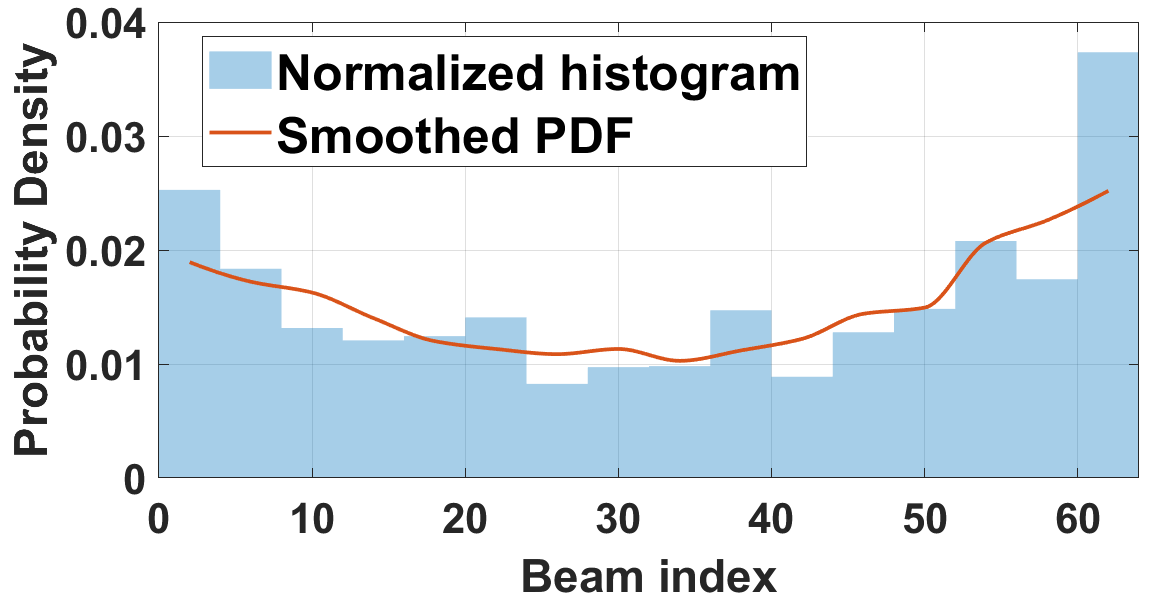}}
        \subfigure[Scenario 31.]
    {\label{fig:S31_beam_density} \includegraphics[width=0.235\textwidth]{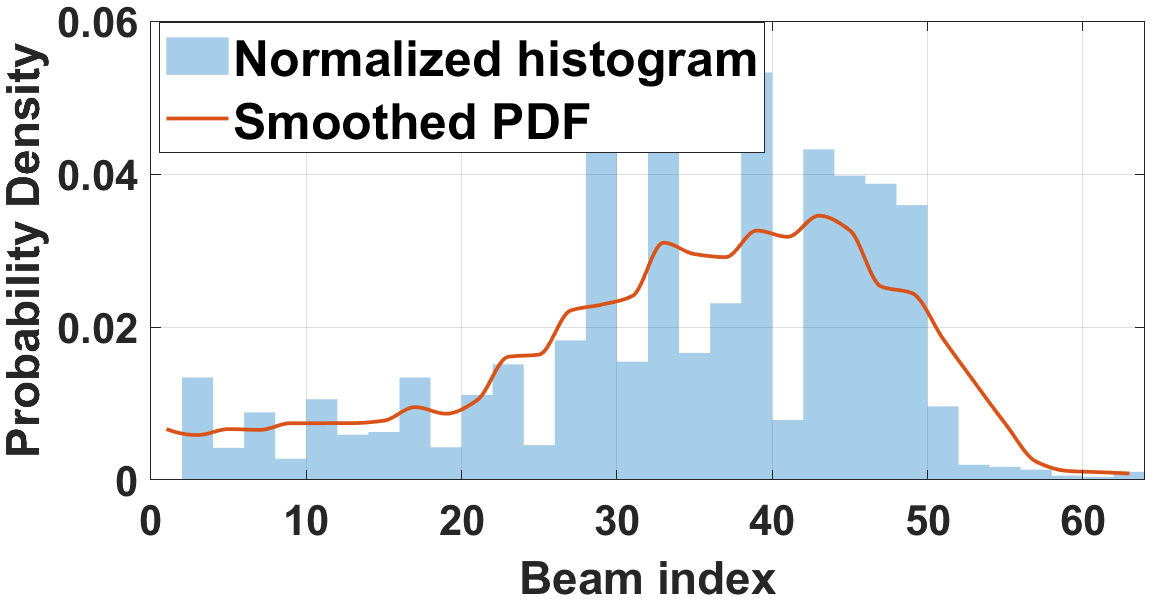}}
    \vspace{-2mm}
\caption{Probability density distributions of the optimal beam index in DeepSense 6G Scenarios 9 and 31. The bars represent the normalized empirical histograms, and the curves show the corresponding smoothed probability density estimates.}
    \label{fig:pdf_beam_index}
\end{figure}

In this section, we evaluate the proposed ML model for cross-environment beam tracking using scenarios 9 and 31 of the DeepSense 6G dataset~\cite{alkhateeb2023deepsense}, which provides real-world sensory data and optimal beam labels for mmWave communications. Figs.~\ref{fig:sceanrio_illustration} and~\ref{fig:pdf_beam_index} present the environmental layouts and the optimal beam-index distributions of the two considered DeepSense 6G scenarios. Scenarios~9 and~31 have distinct BS locations, sensor fields of view, surrounding structures, and traffic patterns, which lead to scenario-dependent propagation and mobility characteristics, as reflected by the empirical beam-index distributions in Fig.~\ref{fig:pdf_beam_index}. In particular, the optimal beam indice in scenario~9 exhibits a broader distribution with high density at both low and high values, whereas those of scenario~31 mainly concentrates in the middle beam-index range. These observations reveal clear differences between the two scenarios in terms of environmental geometry, UE mobility, and interference conditions. Consequently, a performance evaluation involving both scenarios provides a practical assessment for the robustness and generalizability of the proposed lightweight model. 

{\bf Dataset Preparation:} The datasets from scenarios 9 and 31 are merged for training and testing. In particular, $80\%$ of the samples from each scenario are combined to form the training set, while the remaining $20\%$ form the testing set. This ensures that both scenarios are equally represented during both training and testing. All models are trained on the merged training set and evaluated on the merged testing set. In the considered testbed, the camera is deployed close to the BS antenna array with a fixed relative geometry, and the beam labels are collected under the same setup. Therefore, the mapping from visual observations to antenna beam indices is implicitly captured by the supervised training process.


{\bf Training Setup:}  For each time step $t$, we use the past $L=8$ observations to predict the optimal beams for current and future $J=3$ time slots. Both scenarios use a codebook with $64$ candidate beams. During training, we use the proposed data augmentation strategy with $p_{\mathrm{morph}}=0.6$, $p_{\mathrm{blur}}=0.4$, $p_{\mathrm{noise}}=0.5$, and $\sigma_{\rm image}=0.01$. The focal loss parameters are set to $\alpha=1$ and $\gamma=2$, and the temperature for label smoothing is set to $T=2$. The learning rate and loss weight $\lambda$ are selected by grid search. Further implementation details are provided in the released source code\footnote{The source code is available at \url{https://github.com/WillysMa/Lightweight-Beam-Tracking-ISAC.git}.}

{\bf Evaluation Metrics:} We use the Top-$k$ accuracy and distance-based accuracy (DBA) \cite{ma2025knowledge} for performance evaluation. The Top-$k$ accuracy measures whether the ground-truth label is among the model's Top-$k$ predicted labels. The DBA metric assigns a score based on the distance between the predicted and ground-truth beams, and is computed using the Top-$3$ accuracy. Note that both the Top-$k$ and DBA scores target one time slot. To reflect the overall performance across all $J+1$ time slots, we also compute the average Top-$k$ (ATop-$k$) and average DBA (ADBA).

\textbf{Baseline Models:}
To provide a strong performance reference, we consider two benchmarks: (1) \textbf{ResNet-GRU:} a high-capacity baseline model employing  pretrained ResNet-18 backbones~\cite{he2016deep} for image encoding, followed by a two-layer GRU and an MHA module for temporal modeling. This baseline prioritizes predictive accuracy at the cost of high model complexity; (2) \textbf{CNN-GRU:} adopts conventional CNNs for image feature extraction that were developed specifically for scenario 9~\cite{ma2025attention}. 

\begin{table}[t]
\centering
\caption{Overall generalization performance of the considered models.}
\label{tab:model_ablation_results}
\setlength{\tabcolsep}{2pt}
\begin{tabular}{c| c c |c c c c}
\hline
Model & \makecell{Data\\augmentation} & \makecell{Label\\smoothing} & ATop-$1$ & ATop-$3$ & ATop-$5$ & ADBA \\
\hline
\multirow{4}{*}{ResNet-GRU}
& $\times$      & $\times$   & 0.3814  & 0.7187 & 0.8546 & 0.8565 \\
& $\checkmark$ & $\times$    & 0.3958 & 0.7371 & 0.8602 & \textbf{0.8706} \\
& $\times$      & $\checkmark$& 0.4036 & 0.7404 & 0.8649 & 0.8639 \\
& $\checkmark$ & $\checkmark$& \textbf{0.4169} & \textbf{0.7565} & \textbf{0.8688} & 0.8655 \\
\hline
\multirow{4}{*}{Proposed}
& $\times$      & $\times$  & 0.3376    & 0.6454 & 0.7743 & 0.7954 \\
& $\checkmark$ & $\times$  &  0.3453   & 0.6696 & 0.7959 & 0.8141 \\
& $\times$      & $\checkmark$& 0.3566 & 0.6810 & 0.8145 & 0.8214 \\
& $\checkmark$ & $\checkmark$& \textbf{0.3882} & \textbf{0.7301} & \textbf{0.8403} & \textbf{0.8397} \\
\hline
\multirow{4}{*}{CNN-GRU \cite{ma2025attention}}
& $\times$      & $\times$ &  0.3260   & 0.6533 & 0.8086 & 0.8165 \\
& $\checkmark$ & $\times$  &  0.3588  & 0.6736 & 0.7966 & 0.8166 \\
& $\times$      & $\checkmark$& \textbf{0.3789} & \textbf{0.7036} & \textbf{0.8268} & \textbf{0.8325} \\
& $\checkmark$ & $\checkmark$& 0.3746 & 0.6887 & 0.8152 & 0.8274 \\
\hline
\end{tabular}
\end{table}

\begin{table}[t]
\vspace{3mm}
\centering
\caption{Generalization performance of the best models for $J+1$ time slots.}
\label{tab:best_model_timeslot}
\setlength{\tabcolsep}{4pt}
\begin{tabular}{c| c |c c c c}
\hline
Model & Metric & $t_1$ & $t_2$ & $t_3$ & $t_4$ \\
\hline
\multirow{4}{*}{ResNet-GRU}
& Top-$1$ & 0.4198 & 0.4189 & 0.4208 & 0.4081 \\
 & Top-$3$ & 0.7578 & 0.7611 & 0.7578 & 0.7494 \\
 & Top-$5$ & 0.8696 & 0.8714 & 0.8677 & 0.8663 \\
 & DBA     & 0.8643 & 0.8686 & 0.8666 & 0.8626 \\
\hline
\multirow{4}{*}{Proposed}
& Top-$1$ & 0.3913 & 0.3927 & 0.3862 & 0.3824 \\
 & Top-$3$ & 0.7368 & 0.7331 & 0.7349 & 0.7158 \\
 & Top-$5$ & 0.8415 & 0.8415 & 0.8410 & 0.8373 \\
 & DBA     & 0.8415 & 0.8423 & 0.8403 & 0.8346 \\
\hline
\multirow{4}{*}{CNN-GRU \cite{ma2025attention}}
& Top-$1$ & 0.3791 & 0.3843 & 0.3829 & 0.3693 \\
 & Top-$3$ & 0.7073 & 0.7101 & 0.7008 & 0.6961 \\
 & Top-$5$ & 0.8256 & 0.8336 & 0.8294 & 0.8186 \\
 & DBA     & 0.8331 & 0.8366 & 0.8316 & 0.8285 \\
\hline
\end{tabular}
\end{table}


\begin{figure}[t]
\vspace{-2mm}
\small
    \centering
    \hspace{-5mm}
        \subfigure[Model complexity]
    {\label{fig:complexity_models} \includegraphics[width=0.24\textwidth]{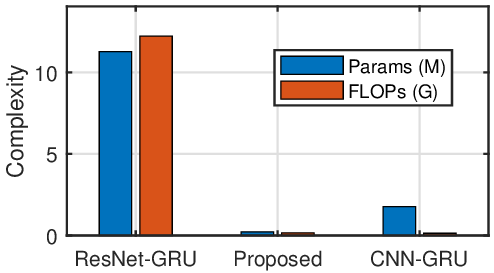}}
        \subfigure[Generalization performance]
    {\label{fig:Score_models} \includegraphics[width=0.24\textwidth]{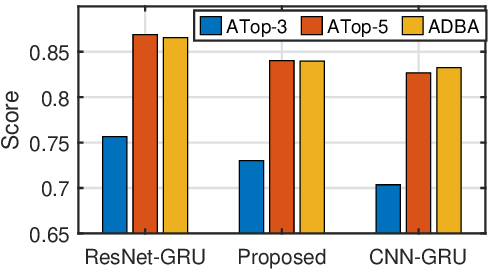}}
    \vspace{-2mm}
\caption{Comparison of model complexity and generalization performance.}
    \label{fig:comp_acc_comparison}
\end{figure}


Table~\ref{tab:model_ablation_results} summarizes the overall generalization performance averaged across the four prediction time slots. For ResNet-GRU, the best ATop-$1$, ATop-$3$, and ATop-$5$ are achieved when both data augmentation and label smoothing are applied, reaching $41.69\%$, $75.65\%$, and $86.88\%$, respectively. The proposed lightweight model also benefits significantly from the two training strategies, improving ATop-$1$ from $33.76\%$ to $38.82\%$, ATop-$3$ from $64.54\%$ to $73.01\%$, ATop-$5$ from $77.43\%$ to $84.03\%$, and ADBA from $79.54\%$ to $83.97\%$. These results show that the proposed augmentation and soft supervision strategies improve model's generalization ability, especially for compact models with limited representation capacity. In particular, label smoothing contributes more to beam prediction performance than data augmentation. The considered model consistently performs better with label smoothing alone than with data augmentation alone while the CNN-GRU model achieves its best performance using only label smoothing. Compared with CNN-GRU~\cite{ma2025attention}, the proposed model achieves higher ATop-$1$, ATop-$3$, ATop-$5$, and ADBA under the best setting, demonstrating the effectiveness of the proposed lightweight architecture.
 
Table~\ref{tab:best_model_timeslot} further reports the temporal performance of the best configuration for each model. ResNet-GRU achieves the highest accuracy across all prediction horizons, while the proposed model consistently outperforms CNN-GRU in terms of Top-$1$, Top-$3$, Top-$5$, and DBA scores. The prediction accuracy for the later time slot $t_4$ is slightly lower than those at $t_1-t_3$, reflecting the increased uncertainty and difficulty for long-term beam tracking. Despite this, the proposed model achieves above $83.5\%$ Top-$5$ accuracy and attains a DBA score of around $84\%$ for all four time slots. This indicates that the proposed model can maintain robust prediction performance for both current and future beam prediction, despite its substantially reduced model size.

Fig.~\ref{fig:comp_acc_comparison} compares the model complexity and the corresponding best generalization performance. Although ResNet-GRU achieves the best prediction accuracy, it requires $11.3$M parameters and $12.2$G FLOPs, making it computationally expensive. In contrast, the proposed model contains only $0.217$M parameters and requires $155$M FLOPs, corresponding to reductions in the number of parameters and FLOPs by factors of approximately $52$ and $79$, respectively, compared with ResNet-GRU. Compared with CNN-GRU~\cite{ma2025attention}, the proposed model achieves higher prediction accuracy with an approximately $8$-fold reduction in the number of parameters. These results show that the proposed design provides a favorable tradeoff between prediction accuracy and model complexity for cross-environment long-term beam tracking.

\section{Conclusions}
This paper proposed a vision-aided lightweight beam tracking framework in cross-environment mmWave communications. Using real-world images from two geometrically distinct DeepSense~6G scenarios, we formulated long-term beam tracking as a Seq2Seq classification problem and developed a compact model based on depthwise separable convolutions. Hierarchical data augmentation and beam power-based label smoothing were incorporated into the training process to improve model robustness and generalization ability. Simulation results show that the proposed model achieves about $84\%$ beam prediction accuracy for the current and three future time slots, outperforming the SOTA CNN-GRU solution while reducing the number of parameters and FLOPs by factors of approximately $52$ and $79$, respectively, compared with the high-capacity ResNet-GRU baseline. These results demonstrate the potential of lightweight sensing-aided models with suitable regularized learning strategies for maintaining satisfactory beam tracking performance under diverse environments. Future work may extend the proposed framework to multimodal sensing and further investigate domain generalization to unseen scenarios.

\bibliographystyle{IEEEtran}
\bibliography{conf_short,jour_short,refs-my,own}

\end{document}